\documentclass[twocolumn,amsmath,amssymb]{revtex4}
\usepackage{graphicx}
\usepackage{dcolumn}
\usepackage{epsfig}
\usepackage{bm}

\begin{document}
\title{Robustness of multiparty nonlocality to local decoherence}

\author{Sung Soon Jang}
\affiliation{IT Convergence Technology Research Division,
Electronics and Telecommunications Research Institute, Daejeon
305-701, Korea}
\author{Yong Wook Cheong}
\affiliation{Quantum Photonic Science Research Center, Hanyang
University, Seoul 133-791, Korea}
\author{Jaewan Kim}
\affiliation{School of Computational Sciences, Korea Institute for
Advanced Study, Seoul 130-722, Korea}
\author{Hai-Woong Lee}
\affiliation{Department of Physics, Korea Advanced Institute of
Science and Technology, Daejeon 305-701, Korea}

\date{\today}

\begin{abstract}
We investigate the robustness of multiparty nonlocality under local
decoherence, acting independently and equally on each subsystems. To
be specific, we consider an $N$-qubit GHZ state under
depolarization, dephasing, or dissipation channel, and tested the
nonlocality by violation of Mermin-Klyshko inequality, which is one
of Bell's inequalities for multi-qubit systems. The results show
that the robustness of nonlocality increases with the number of
qubits, and that the nonlocality of an $N$-qubit GHZ state with even
$N$ is extremely persistent against dephasing.
\end{abstract}

\maketitle

\section{Introduction}

Nonlocality is one of the most surprising and important features of
quantum physics. It is firstly referred by Einstein, Podolsky, and
Rosen, and has been receiving an enormous attention and interest of
scientists since Bell\cite{Bell64,CHSH69} designed an inequality to
expose it. Now, it is an essential part of quantum information
science and quantum physics foundation.

However, nonlocality, like other quantum features, is easily
destroyed by decoherence, and the concrete pictures of its
destruction is not well known. One might gain insight into it by
analyzing nonlocal states under decoherence. Werner\cite{Werner89}
studied a maximally entangled two qubits under white noise, and
Kazlikowski et. al.\cite{two-party} extended Werner's work to a two
$d$-dimensional system.

In particular, we are interested in the robustness of nonlocality
according to the number of consisting particles. Nonlocality is
hardly seen at macroscopic level because of strong decoherence,
caused by many interactions with environment. In a previous
study\cite{SiKe02}, it was shown that entanglement of an $N$-qubit
GHZ state is more robust to depolarization as the number of qubits
increases. The authors examined persistency of inseparability, which
is the mathematical definition of entanglement. In this paper, we
are interested in persistency of nonlocality, which is physically
meaningful nature of entanglement.

We examine the decoherence process under realistic noise. Besides
depolarization we consider two other common decoherence processes,
dissipation and dephasing, which are called $T_1$ and $T_2$ process,
respectively. For simplicity, we assume equal and independent local
decoherence on each sub-particles. Cabello and Feito\cite{CaFe05}
studied Bell's inequality with depolarization and dephasing, and
found that violations of Bell's inequality for a two qubit system
are extremely robust to dephasing. Although they extended their work
to three and four qubits, they failed to show this extreme
robustness, because they searched the violations of Bell's
inequality over limited Hilbert space.

In this paper, we study the nonlocality of multi-qubit GHZ states
under three decoherences. The paper is organized as follows. Section
II.A introduces our nonlocality criteria, Mermin-Klyshoko(MK)
inequality, section II.B describes three decoherence models and
decohered GHZ states, and section II.C construct the inequalities of
the states. In section III, we numerically test the inequalities and
discuss the results. Finally, section IV summarizes the conclusions.

\section{Methods}

\begin{figure}
\includegraphics[width=8cm]{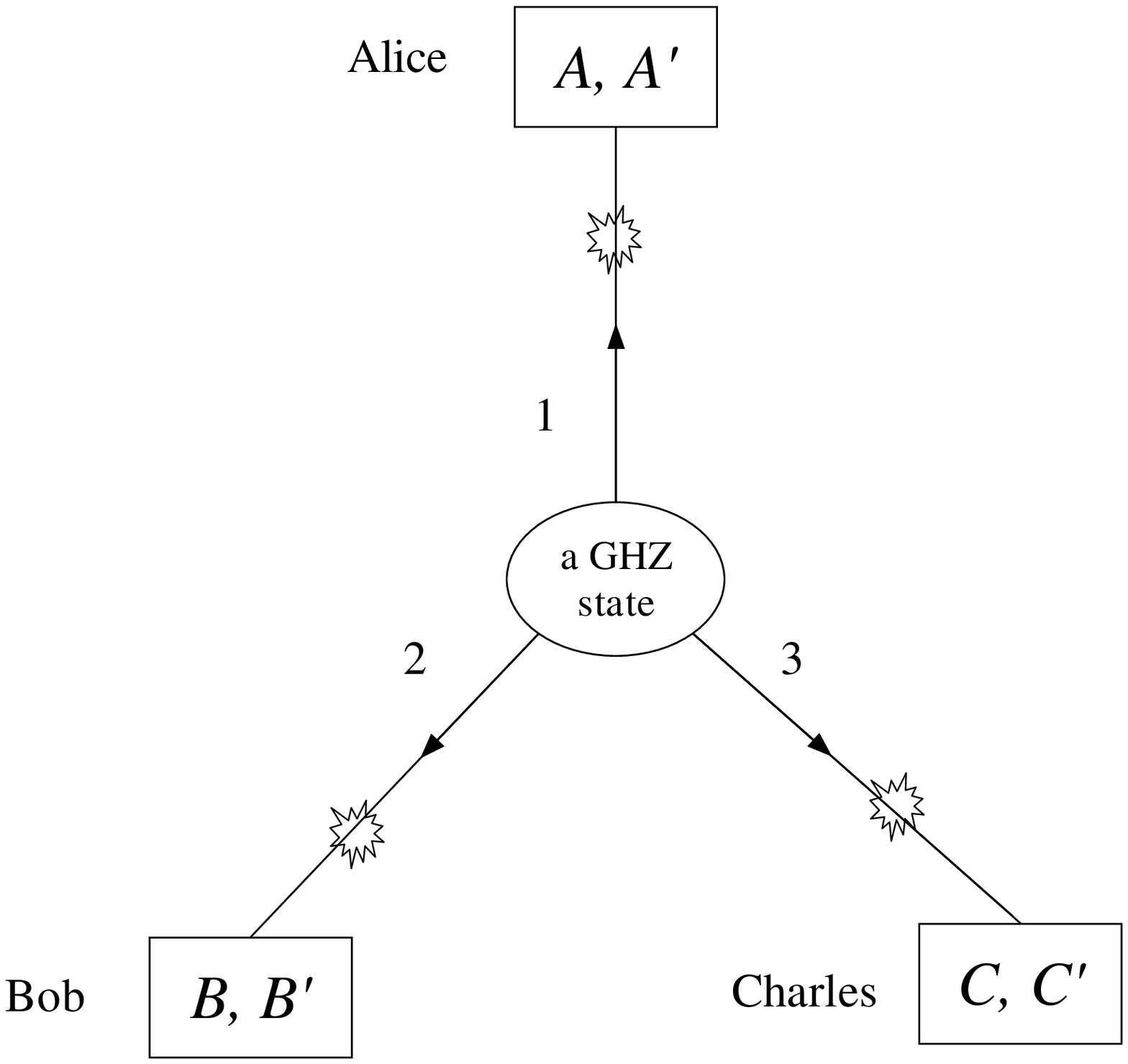} \caption{Experimental
Scheme. The source emits a GHZ state $ \frac{1}{2} (|000\rangle +
|111\rangle )$, where equal and independent local decoherence happens to
each individual qubit. The decoherence is represented by flash
symbols in the figure. To estimate nonlocality,
Alice performs a projection measurement $A$
or $A'$ on her qubit $1$, Bob performs a measurement $B$ or $B'$,
and Charles performs a measurement $C$ or $C'$, respectively.}
\label{fig1}
\end{figure}

We assume the following situation. Parties are sharing multi-qubit
GHZ states, where individual qubit suffers equal and independent
local decoherence. To estimate nonlocality, parties measure their
qubits in one of two promised measurements and consult the results.
Fig.~1 shows the case where the number of party is three. We will
use the term ``a GHZ state'' not only for a three qubit state but
also for a multi-qubit GHZ state.

\subsection{How to know nonlocal or not?}

We select Mermin-Klyshko(MK)\cite{MK} inequalities as nonlocality
tester. We want to test the nonlocality of symmetric $N$-qubit
states by dichotomic measurements(two different measurements per
party). In that situations, the MK inequality has many advantages;
it is simpler than other Bell's inequalities for multi-qubit
systems\cite{Generic-Bell}, invariant under permutation, and
maximally violated by a GHZ state. However, the inequality also has
a limitation; for some nonlocal states, it cannot determine the
nonlocality.

MK inequality of an $N$-qubit system is written as
\begin{equation}
|\langle {\cal B}_N \rangle_{\psi}| \le 1, \label{MK1}
\end{equation}
where, the operator ${\cal B}_N$ is an $N$-qubit {\it Bell
operator}, and a {\it Bell value} $\langle {\cal B}_N
\rangle_{\psi}$ is the expectation value of it, given a state
$\psi$. Although the bound is $1$ as seen in Eq. (\ref{MK1}), the
Bell values of MK inequalities can rise to $2^{(N-1)/2}$,  quantum
mechanically. The Bell operator is recursively defined as follows.
\begin{equation}
{\cal B}_N \equiv \frac{1}{2}{\cal B}_{N-1}  \otimes \left( {K + K'}
\right) + \frac{1}{2}{\cal B}'_{N-1}  \otimes \left( {K - K'}
\right), \label{MK2}
\end{equation}
where the symbol $\otimes$ is vector product. The capital letters
stand for individual party's measurements(observables), and $i$th
party's two measurements are presented by the primed and non-primed
$i$th capital ($A,A',B,B',\cdots,K,K'$). The letter $A$(also the
lower-case $a$) refers to first party and the letter $K$(also $k$)
refers to $N$th party. The primed Bell operator ${\cal B}'_N$ is
obtained from ${\cal B}_N$ by exchanging every primed and non-primed
capitals. The starting of recursion is ${\cal B}_2$, which is a CHSH
inequality, Eq. (\ref{B2}) below. For example, the Bell operators of
two, three, and four MK inequality are the following.
\begin{equation}
{\cal B}_2 = \frac{1}{2}\left( AB + AB' + A'B - A'B' \right)
\label{B2}
\end{equation}
\begin{equation}
{\cal B}_3 = \frac{1}{2}\left( ABC' + AB'C + A'BC - A'B'C' \right)
\label{B3}
\end{equation}
\begin{equation}
\begin{array}{l}
{\cal B}_4 =
 \frac{1}{4}(-ABCD+ABCD'+ABC'D+ABC'D' \\
 \quad \quad \quad +AB'CD+AB'CD'+AB'C'D-AB'C'D'  \\
 \quad \quad \quad +A'BCD+A'BCD'+A'BC'D-A'BC'D'  \\
 \quad \quad \quad  +A'B'CD-A'B'CD'-A'B'C'D-A'B'C'D')
\end{array}
\label{B4}
\end{equation}

\subsection{Decoherence Models}

Depolarization, dephasing, and dissipation on a qubit are our
decoherence models. Let's define $p$ as {\em the degree of
decoherence of an individual qubit}, which lies between $0$ and $1$,
where the value $0$ means no decoherence, and $1$ means complete
decoherence. The depolarization process to a state with a
decoherence degree $p$ is represented by
\begin{equation}
\left| i \right\rangle \left\langle j \right| \to \left( {1 - p}
\right)\left| i \right\rangle \left\langle j \right| + p\delta _{ij}
\frac{I}{2}. \label{depol}
\end{equation}
The dephasing process is described by
\begin{equation}
\left| i \right\rangle \left\langle j \right| \to \left( {1 - p}
\right)\left| i \right\rangle \left\langle j \right| + p\delta _{ij}
\left| i \right\rangle \left\langle j \right|. \label{dephase}
\end{equation}
The dissipation is the process losing energy, and thus it changes a
state to a specific state (e.g., a ground state). We choose
$|0\rangle$ as the ground state. The dissipation process is then
described by
\begin{equation}
\begin{array}{l}
 \left| i \right\rangle \left\langle i \right| \to \left( {1 - p} \right)\left| i \right\rangle \left\langle i \right| + p\left| 0 \right\rangle \left\langle 0 \right| \\
 \left| i \right\rangle \left\langle j \right| \to \left( {1 - p} \right)^{1/2} \left| i \right\rangle \left\langle j \right|\quad \quad \quad {\rm where} \quad i \ne j. \\
 \end{array} \label{dissip}
\end{equation}
These three decoherences can be mathematically simplified to two
processes, population transfer and dephasing.

An initial GHZ state suffers one of the above decoherence processes,
which act equally and independently on each qubits. If every
individual qubit of a GHZ state is partially depolarized as
Eq.~(\ref{depol}), then the density matrix of the depolarized GHZ
state becomes
\begin{equation}
\begin{array}{l}
\displaystyle\frac{1}{2} [ \left( {\left( {1 - {\textstyle{p \over
2}}} \right)\left| 0 \right\rangle \left\langle 0 \right| +
{\textstyle{p \over 2}}\left| 1 \right\rangle \left\langle 1
\right|} \right)^{ \otimes N}
 + \left( {\textstyle{p \over
2}}\left| 0 \right\rangle \left\langle 0 \right| \right. + \\
\left. \left( {1 - {\textstyle{p \over 2}}} \right)\left| 1
\right\rangle \left\langle 1 \right| \right)^{ \otimes N}
 + \left( {1 - p} \right)^N ( {\left| 0
\right\rangle \left\langle 1 \right|^{ \otimes N}  + \left| 1
\right\rangle \left\langle 0 \right|^{ \otimes N} } ) ],
\label{depol-GHZ}
\end{array}
\end{equation}
where $ |i\rangle\langle j|^{\otimes N} \equiv |ii\cdots
i\rangle\langle jj\cdots j|$. The partially dephased GHZ state by an
amount $p$ is
\begin{equation}
\displaystyle\frac{1}{2} [ \left| 0 \right\rangle \left\langle 0
\right|^{ \otimes N}  + \left| 1 \right\rangle \left\langle 1
\right|^{ \otimes N} + \left( {1 - p} \right)^N ( {\left| 0
\right\rangle \left\langle 1 \right|^{ \otimes N} + \left| 1
\right\rangle \left\langle 0 \right|^{ \otimes N} })],
\label{dephase-GHZ}
\end{equation}
and the partially dissipated GHZ state is
\begin{equation}
\begin{array}{l}
\displaystyle\frac{1}{2} [ \left| 0 \right\rangle \left\langle 0
\right|^{ \otimes N}  + \left( {p\left| 0 \right\rangle
\left\langle 0 \right| + \left( {1 - p} \right)\left| 1
\right\rangle \left\langle 1 \right|} \right)^{
\otimes N}  \\
+ \left( {1 - p} \right)^{N/2}( {\left| 0 \right\rangle
\left\langle 1 \right|^{ \otimes N}  + \left| 1 \right\rangle
\left\langle 0 \right|^{ \otimes N} }) ]. \label{dissip-GHZ}
\end{array}
\end{equation}

\subsection{Correlations and MK inequalities}

Every projection measurement over a qubit can be represented by a
unit vector in the Bloch sphere. An arbitrary observable $A$ can be
written as
\begin{equation}
A = \left( {\sigma _x \cos \phi _a  + \sigma _y \sin \phi _a }
\right)\sin \theta _a  + \sigma _z \cos \theta _a,
\end{equation}
$\sigma_x$, $\sigma_y$, and $\sigma_z$ are pauli spin matrices, and
$\theta_a$ and $\phi_a$  are the polar and azimuthal angles
corresponding to $A$ in the Bloch sphere representation. An
eigenvalue of the observable(i.e. a measurement result) is $1$ or
$-1$.

\begin{widetext}
The measurements conducted by the participating parties construct
{\em correlation}, which is the ensemble average of the
multiplication of every party's observable $\langle AB \cdots K
\rangle \equiv Tr\{AB\cdots K \rho \}$. The correlation of a GHZ
state is
\begin{equation}
\langle AB \cdots K \rangle_{GHZ} = \left[ {\frac{{1 + \left( { - 1}
\right)^N }}{2}\cos \theta _a \cdots \cos \theta _k  + \cos \left(
{\phi _a  +  \cdots  + \phi _k } \right)\sin \theta _a \cdots \sin
\theta _k } \right], \label{correl-GHZ}
\end{equation}
the correlation of a depolarized GHZ state (\ref{depol-GHZ}) is
\begin{equation}
\left\langle {AB \cdots K} \right\rangle _{depol}  = \left( {1 - p}
\right)^N \left[ {\frac{{1 + \left( { - 1} \right)^N }}{2}\cos
\theta _a  \cdots \cos \theta _k  + \cos \left( {\phi _a  + \cdots +
\phi _k } \right)\sin \theta _a  \cdots \sin \theta _k } \right],
\label{correl-depol}
\end{equation}
the correlation of a dephased GHZ state (\ref{dephase-GHZ}) is
\begin{equation}
\left\langle {AB \cdots K} \right\rangle _{dephase}  = \left[
{\frac{{1 + \left( { - 1} \right)^N }}{2}\cos \theta _a  \cdots \cos
\theta _k  + \left( {1 - p} \right)^N \cos \left( {\phi _a  + \cdots
+ \phi _k } \right)\sin \theta _a  \cdots \sin \theta _k } \right],
\label{correl-dephase}
\end{equation}
and the correlation of a dissipated GHZ state (\ref{dissip-GHZ}) is
\begin{equation}
\left\langle {AB \cdots K} \right\rangle _{dissip}  = \left[
{\frac{{1 + \left( {2p - 1} \right)^N }}{2}\cos \theta _a  \cdots
\cos \theta _k  + \left( {1 - p} \right)^{N/2} \cos \left( {\phi
_a + \cdots  + \phi _k } \right)\sin \theta _a  \cdots \sin \theta
_k } \right]. \label{correl-dissip}
\end{equation}
\end{widetext}
Since MK inequalities are sum of correlations, as seen in
(\ref{B2})-(\ref{B4}), the inequalities of a pure GHZ state and
those of the state under depolarization(under dephasing or
dissipation) can be constructed from (\ref{correl-GHZ}) and
(\ref{correl-depol})(Eq. (\ref{correl-dephase}) or Eq.
(\ref{correl-dissip})), respectively.

Comparing the correlations Eq.
(\ref{correl-GHZ}),(\ref{correl-depol}) and (\ref{correl-dephase}),
we observe that
\begin{equation}
\begin{array}{lcl}
\langle A B \cdots K\rangle_{depol} & = & (1-p)^N \langle A
B\cdots K \rangle_{GHZ} \\
\langle A B \cdots K \rangle_{dephase|odd}  & = & (1-p)^N \langle A
B\cdots K \rangle_{GHZ},
\end{array}
\end{equation}
where the last equality holds when the number of qubits is odd.
Therefore, Bell inequalities made from the above correlations
satisfy
\begin{equation}
\langle {\cal B}_N \rangle_{depol} = \langle {\cal B}_N
\rangle_{dephase|odd} = (1-p)^N \langle {\cal B}_N \rangle_{GHZ}.
\label{trivial-MK}
\end{equation}

\section{Results and Discussion}

\begin{figure}
\includegraphics[width=8cm]{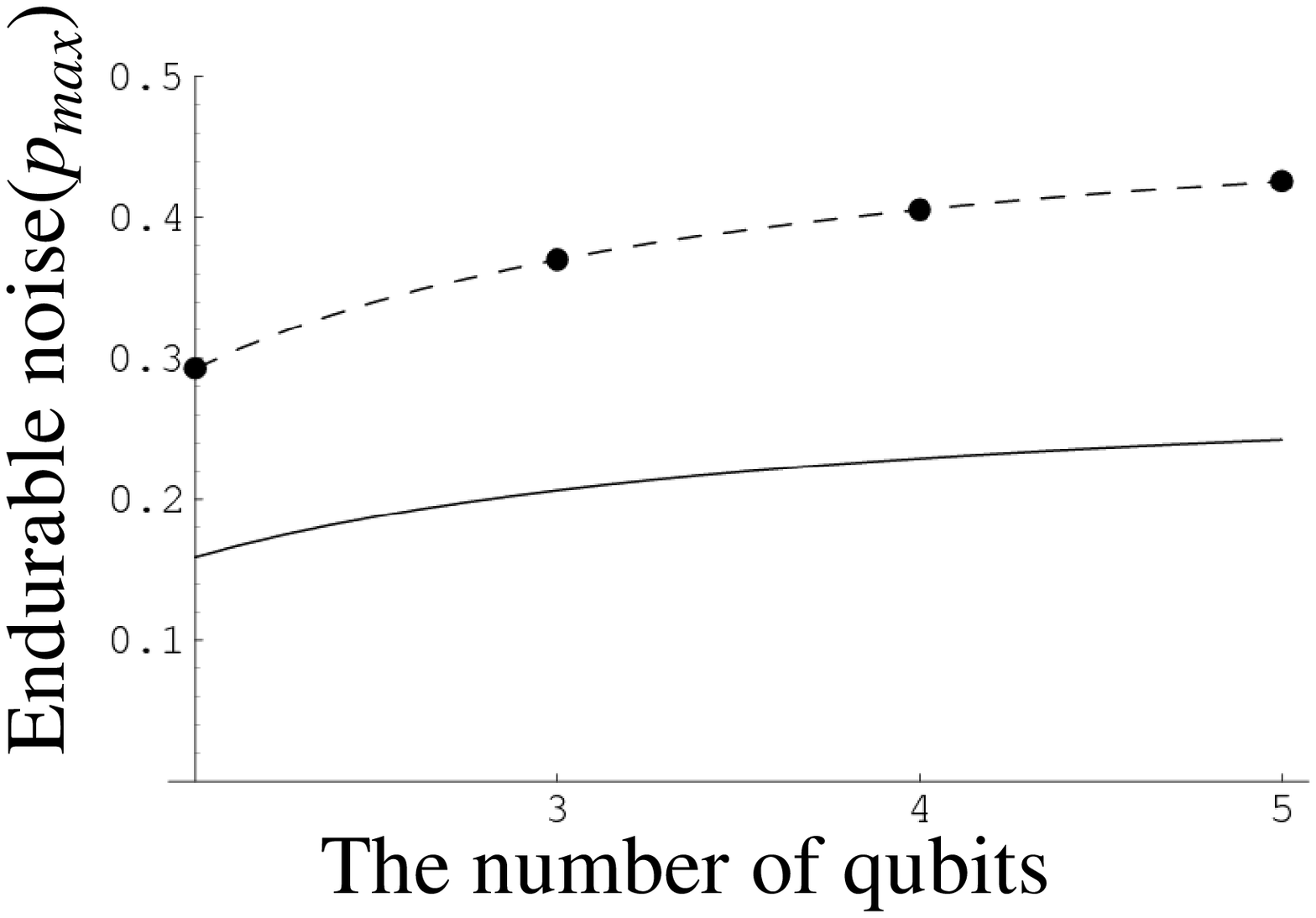} \caption{The noise for
which MK inequality is still violated. The solid line is sustainable
noise for a GHZ state under depolarziatin or the state of odd $N$
under dephasing, Eq. (\ref{pmax1}). The points are for a GHZ state
under dissipation by numerical calculations, and fits well to Eq.
(\ref{pmax2}), which is represented by the dashed line. }
\label{fig2}
\end{figure}

We can easily observe the robustness of multi-qubit nonlocality of a
GHZ state under depolarization and the state of odd $N$ under
dephasing. The robustness is judged by the amount of the maximum
noise for which the Bell inequality is still violated, i.e.
nonlocality prolongs. From Eq. (\ref{trivial-MK}) the violation
condition of the states is
\begin{equation}
(1-p)^N \langle {\cal B}_N \rangle_{GHZ} > 1.
\end{equation}
In the above inequality, the noise $p$ becomes maximal at the
highest Bell value of a GHZ state $\langle {\cal B}_N
\rangle_{GHZ}$, which is $2^{(N-1)/2}$. Consequently, the maximum
noise $p_{max}$ is given by
\begin{equation}
p_{max} = 1 - 2^{(\frac{1}{N} - 1)/2}. \label{pmax1}
\end{equation}
Eq. (\ref{pmax1}) is a monotonically increasing function of N that
asymptotically becomes close to $1-\frac{1}{\sqrt 2}$, and is shown
by the solid line of Fig.~2. In a GHZ state under depolarization and
the state of odd $N$ under dephasing, the nonlocality becomes more
robust as the number of qubits increases.

We numerically investigated violations of MK inequalities for
nontrivial cases: a GHZ state under dissipation and the state of
even N under dephasing. We used MATHEMATICA as a simulation tool,
and evaluated the system up to five qubits.

The nonlocality of a GHZ state also shows persistency against
dissipation with the increasing number of qubits. The maximum noise
$p_{max}$ for which the Bell inequality is still violated is
numerically calculated and is plotted by the points in Fig.~2. The
numerical result fits well to the equation
\begin{equation}
p_{max} = 1 - 2^{\frac{1}{N} - 1}. \label{pmax2}
\end{equation}
Eq. (\ref{pmax2}) is also a monotonically increasing function of N
and is drawn by the dashed line of Fig.~2.

To see how nonlocality changes, we plot the maximum Bell value of a
GHZ state under dissipation in Fig.~3. The curve falls below the
inequality bound by dissipation, and it return to the bound at
complete dissipation, because the fully dissipated state is a ground
state $|00\cdots0\rangle$, which has a classical correlation.

Fig.~4 shows that the nonlocality of a GHZ state with even number of
qubits is extremely robust against dephasing. Dephasing is
considered to be the main cause of destroying quantum coherence.
Although dephasing is very fast in a real system, nonlocality always
survives, if the dephasing is not complete. This effect was already
noted by Cabello and Feito \cite{CaFe05} but not for more than two
qubits. The extreme robustness of a GHZ state under dephasing
critically depends on the parity of the number of qubits. Whether
this parity dependence is real or accidental is not clear yet. To
study this question further we could explore other nonlocality
criteria. We remark that this feature is very closely related to
entanglement purification. One can get a pure GHZ state by consuming
many partially dephased states by the purification, if the dephasing
does not happen completely \cite{MPPVK98}; this is not the case for
depolarization.

\begin{figure}
\includegraphics[width=8cm]{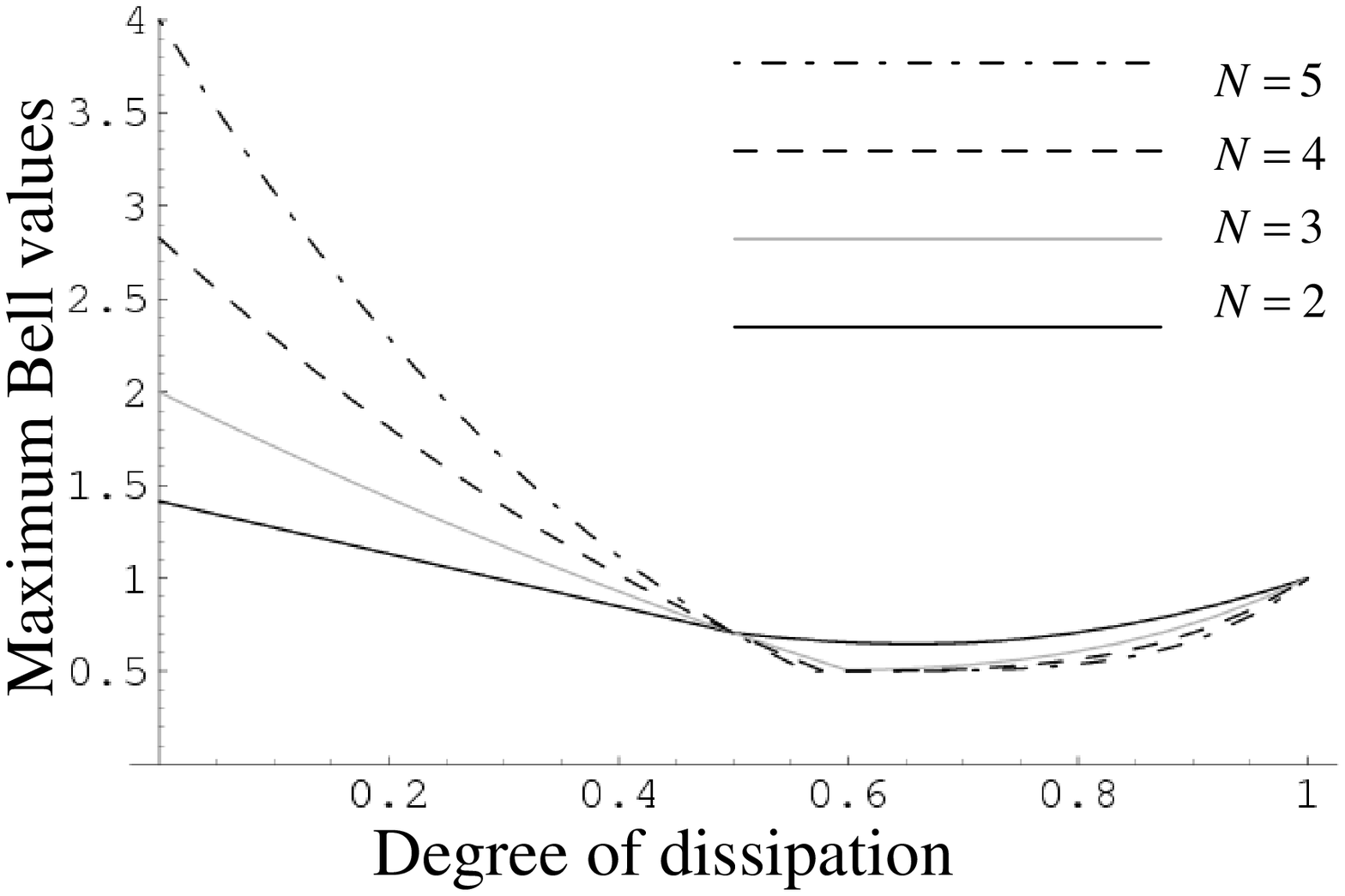} \caption{Maximum Bell value
of the MK inequality as a function of $p$ for states with
dissipation given by eq.~(\ref{dissip}). The number of qubits is
from two to five and is distinguished by line style.} \label{fig3}
\end{figure}

\begin{figure}
\includegraphics[width=8cm]{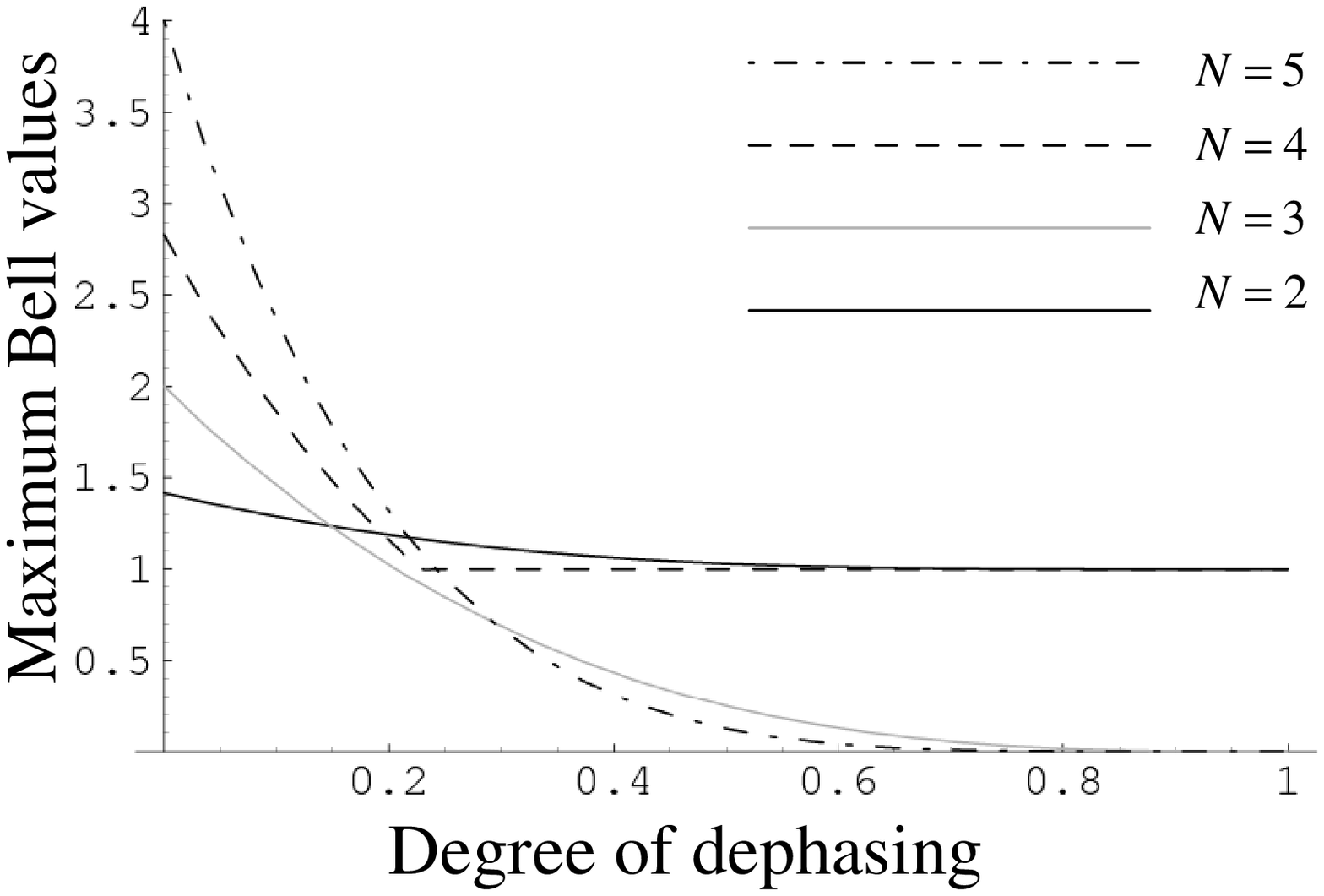} \caption{Maximum Bell value
of the MK inequality as a function of $p$ for states under dephasing
given by eq.~(\ref{dephase}). The number of qubits is from two to
five and is distinguished by line style.} \label{fig4}
\end{figure}

\section{Conclusion}

In summary, we studied the persistency of nonlocality over $N$-qubit
GHZ states under realistic local decoherences: depolarization,
dephasing, and dissipation. We observed nonlocality by testing
violations of MK inequality of decohered GHZ states, using numerical
simulations. This study revealed that multi-qubit GHZ states have
much robust nonlocality with increasing number of qubits, and showed
extreme persistency of a GHZ state of even $N$ against dephasing.
This strong persistency can not be seen for a GHZ state of odd $N$.
Whether this parity dependence is real nature of nonlocality or due
to the incompleteness of nonlocality criterion needs further
investigation. The states, having robust nonlocality, will be very
useful in quantum communication and quantum information with
realistic noisy circumstances.

\begin{acknowledgements}
The authors thank Professor Jinhyoung Lee for helpful discussions.
S. S. Jang and J. W. Kim appreciate the financial support from
the Korea Ministry of Information and Communication. H. W. Lee was
supported by Korea Research Institute of Standard and Science(KRISS).
\end{acknowledgements}

\end{document}